\documentclass[10pt]{iopart}
\usepackage{iopams}  
\usepackage{graphicx}
\usepackage[all]{nowidow}

\begin{document}

\title[Controlling spoke]{Controlling azimuthal spoke modes in cylindrical Hall thruster using a segmented anode}

\author{Yuan Shi$^{1,2}$ and Yevgeny Raitses$^2$}

\address{$^1$ Department of Astrophysical Sciences, Princeton University, Princeton, New Jersey 08544, USA}
\address{$^2$ Princeton Plasma Physics Laboratory, Princeton University, Princeton, New Jersey 08543, USA}

\ead{yshi@pppl.gov, yraitses@pppl.gov}

\vspace{10pt}
\begin{indented}
\item[]August 2017
\end{indented}

\begin{abstract}
Azimuthal spoke-like modes commonly occur in $\mathbf{E}\times\mathbf{B}$ discharges. The natural occurrence of the spoke mode is correlated with changes of plasma parameters. Here, instead of allowing these changes to occur naturally, we report a technique for actively controlling the spoke by adjusting the boundary condition at the anode. The technique is demonstrated using a cylindrical Hall thruster equipped with a segmented anode. By varying the voltage and the relative phases of the anode segments, properties of the azimuthal mode can be altered substantially, as shown by fast camera images and probe diagnostics. This technique may be extended to other Hall discharges in order to either induce or suppress azimuthal activities, and thereby controlling the operation and the performance of the devices.
\end{abstract}

%
\vspace{2pc}
\submitto{\PSST}
%
\maketitle
%
\ioptwocol

\section{Introduction}
Coherently rotating azimuthal modes are commonly seen in Hall plasmas, which can be produced in helicon source \cite{Schroder2004spatial}, Hall thrusters \cite{Parker2010transition,Mcdonald2011rotating}, and Penning discharges \cite{Raitses2015Penning}. 
In these Hall plasmas, where the macroscopic electric field has a component perpendicular to the background magnetic field, spoke-like modes arise in the $\mathbf{E}\times\mathbf{B}$ direction. However, the frequencies of the azimuthal modes are far below what is expected from the $\mathbf{E}\times\mathbf{B}$ drift, and the exact mechanism of the formation of spoke-like structures is still currently under investigations \cite{Sakawa2000Growth,Smolyakov2016fluid,Romadanov2016structure,Boeuf2017tutorial,Carlsson2018PIC}.
Regardless of its mechanism, the occurrence of the spoke mode is known to correlate with performance changes in Hall plasma devices. 
For example, when the spoke appears in cylindrical Hall thrusters (CHT), electron cross field transport suddenly increases \cite{Ellison2012cross}, leading to a decrease of the device efficiency.
On the contrary, in conventional annular Hall thrusters, the spoke is associated with more efficient regimes of operation \cite{Sekerak2015azimuthal}.  
Moreover, performance changes have also been observed in magnetron sputtering devices \cite{Keudell2016control,Anders2017spoke} when azimuthal activities arise.
Therefore, depending on the device configuration, the operation regime, and facility effects, the presence of the spoke may have either positive or negative impact on the performance, which is a primary concern of Hall plasma devices.

The azimuthal activities can be altered by a number of passive techniques. For example, increasing the background gas pressure leads to an increase of the spoke frequency, until the azimuthal mode becomes chaotic \cite{Raitses2010background}. 
In addition, the frequency of the spoke, which scales as $\sqrt{B_0/m_i}$ where $B_0$ is the background magnetic field and $m_i$ is the mass of plasma ions \cite{Raitses2015Penning}, can be tuned by changing the magnetic field and plasma species. 
Moreover, the spoke activities in the main discharge is affected by how the cathode operates. For example, when operating the cathode-neutralizer in the current overrun regime, the spoke in a CHT can be suppressed \cite{Parker2010transition}.
Finally, without changing conditions of the discharge, Griswold $ \it{et. al.}$ manage to suppress the spoke in the CHT using a resistive feedback \cite{Griswold2012feedback}. In their setup, an anode with four azimuthal segments is used, where each segment is connected in series with a resistor. The local increase of the spoke-induced current then results in a potential drop on the anode segment, creating a negative feedback that suppresses the $m=1$ azimuthal mode and reduces the total discharge current.

In this paper, we further exploit the idea that azimuthal mode may be controlled by changing the boundary conditions. Here, instead of passively damping the spoke, we investigate active driving of azimuthal activities by applying periodic rotating voltages on the anode segments \cite{Shi2013driving}. 
The time-dependent response of the plasma under active driving is diagnosed by a fast camera and a suite of probes.
By changing the driving frequency, local properties of the CHT plasma is noticeably altered. 
This paper is organized as follows. In Sec.~\ref{sec:setup}, we describe the experimental facility and methods used to drive and diagnose the spoke. The experimental results are presented in Sec.~\ref{sec:results}. Conclusion and discussion are given in Sec.~\ref{sec:conclusion}, followed by a summary.

\section{Experimental Setup \label{sec:setup}}
All experiments are performed at the small Hall thruster facility (SHTF) at the Princeton Plasma Physics Laboratory (PPPL) using the 2.6 cm CHT described elsewhere \cite{Raitses2001parametric}. 
A background pressure of \mbox{$\approx 5 \times 10^{-5}$ Torr}, measured by an ionization gauge, is maintained by the steady-state balance of xenon flow through the thruster and gas expulsion via a turbo-molecular pumping system. The propellant flows are controlled and measured by two Millipore \mbox{FC-260} controllers, which indicate a flow rate of \mbox{3 SCCM} to the anode and \mbox{1.5 SCCM} to the cathode.
The thruster is operated under the ``direct'' magnetic field configuration, in which the radial component of the field lines points inwards, fixing a natural $\bf{E} \times \bf{B}$ rotation direction, where the electric field points out of the thruster channel. The back and front magnetic coil currents are set at \mbox{2.8 A} and \mbox{1 A}. Half of the front coil is removed to facilitate the insertion of probes through the CHT channel wall.
A hollow cathode-neutralizer is placed outside the thruster, at \mbox{57.5 mm} radially from the thruster centerline and \mbox{20 mm} axially from the channel exit. To provide operational stability, a \mbox{12-A} cathode heater current is applied, together with a \mbox{0.5-A} cathode keeper current.
The cathode and thruster bodies are allowed to float, while the vacuum chamber is maintained at ground. 

\begin{figure}[t]
	\centering
	\includegraphics[width=0.25\textwidth]{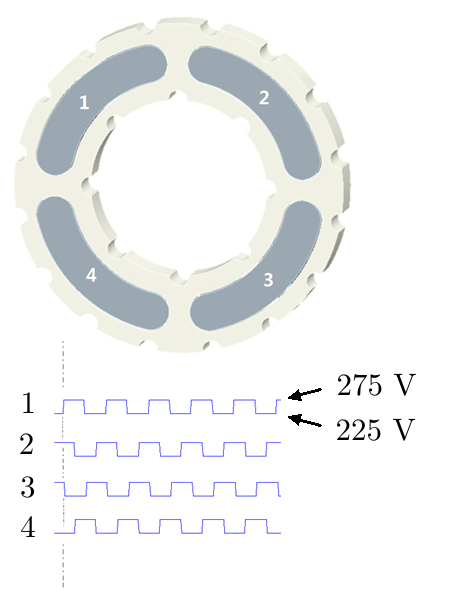}
	\caption{Active control of the azimuthal mode is achieved using an anode cut into four identical azimuthal segments, which are fitted inside an insulating boron nitride ceramics casing. Square-wave voltages with $90^\circ$ phase shifts are applied to consecutive segments to generate an azimuthally rotating potential.}
	\label{fig:design}
\end{figure}

Active control of the azimuthal mode is achieved using an anode cut into four identical azimuthal segments \cite{Ellison2012cross,Griswold2012feedback}. The anode segments are mutually insulated and connected separately to the power supply. 
In the passive operation mode, all anode segments are connected to a constant discharge voltage of 250 V. 
In the active operation mode (Fig.~\ref{fig:design}), each segment is connected to either a \mbox{225 V} or a \mbox{275 V} voltage source through a transistor switch. The four transistor switches are controlled by a logic circuit. The circuit receives clock signal from a single function generator and outputs four square wave signals with $90^\circ$ phase shifts. 
The voltage on the $n$-th segment is then $V_n(t) = 225 + 50 S(t\pm nT/4)$, where $S(t)=\textrm{sgn}(\sin 2\pi t/T)$ is a square wave with frequency $f=1/T$. Notice that the spatially average discharge voltage is 250 V, which is a constant in time regardless of the driving frequency.
The driving frequency $f$ is scanned between \mbox{$\pm 16$ kHz}, where the natural spoke mode, which occurs at the aforementioned operation conditions, has frequency \mbox{$f\approx 6$ kHz}. The positive frequency $f>0$ corresponds to driving in the $\bf{E} \times \bf{B}$ direction, whereas the negative frequency $f<0$ corresponds to driving in the opposite direction.

To monitor the plasma response, a Phantom v7.3 Fast Frame camera is used to visually record the axially integrated light emission from the CHT plasma. The camera is mounted outside of the vacuum chamber and is optically aligned with the plume using a set of mirrors, looking straight into the thruster channel. The light is unfiltered, and the images are captured at a rate of $\sim142,000$ fps with a digital resolution of 64 $\times$ 64 pixels. The integrated pixel intensity across each quadrant, corresponding to each segment, is then used in post processing. 

Langmuir probes, emissive probes, and a plume probe are used to diagnose the time-dependent plasma. The probe systems are similar to what we have used previously \cite{Smirnov2004plasma,Diamant2010ionization}.
Each Langmuir probe is consisted of a single-bored alumina tube with an outer diameter of \mbox{1.14 mm}. The alumina tube is tightly filled with a tungsten rod of \mbox{0.75 mm} in diameter, whose planar tip serves as the probe surface. The Langmuir probe is inserted radially through the thruster body, with the probe tip placed flush with the inner surface of the thruster channel to minimize perturbations to the plasma. 
One sets of Langmuir probes are installed in the near-anode region, at \mbox{$5$ mm} above the centers of the anode segments, in order to take measurements in the ionization zone. Along the same azimuthal directions but in the near-exit region, another set of Langmuir probes are installed at \mbox{$17.5$ mm} from the anode to take measurements in the acceleration zone of the thruster. 
The Langmuir probes are biased to measure the ion saturation currents, from which the plasma density can be inferred.  
To acquire the electron temperature ($T_e$) and plasma potential ($\phi$), two sets of emissive probes are placed similarly in both the ionization and acceleration regions of the thruster. 
By recording the hot ($V_h$) and cold ($V_c$) floating potentials, $T_e$ and $\phi$ are inferred assuming $V_h = \phi - 1.5T_e$ and $V_c = \phi - 5.77T_e$ \cite{Sheehan2016emissive}.
Each emissive probe is constructed using a double-bored alumina tube with one \mbox{0.125 mm} thoriated tungsten filament wire, which is fixed in place by six tungsten filler wires in each hole. The probe is placed such that the filament is ejected into the plasma with the ceramic tube slightly below the CHT channel wall.
Finally, plume measurements of the ejected xenon ions are obtained using a 1.1-cm diameter plume probe.
The plume probe is constituted of two Langmuir probes. One Langmuir probe, with a guarding sleeve, faces the plume and measures the upstream ion saturation current. The other Langmuir probe faces backward and measures the background ion saturation current. The background-subtracted plume current is then obtained when these two Langmuir probes are used in conjunction.
The plume probe is biased at -30 V in order to reach the ion saturation regime. The radial distance between the plume probe and the center of the thruster exit is 15 cm. 
The plume probe is mounted on a rotating arm, which can be moved by a motor to scan ion fluxes at each angular location around the thruster exit. 

\section{Experimental results \label{sec:results}}

\begin{figure*}[t]
	\centering
	\includegraphics[width=0.8\textwidth]{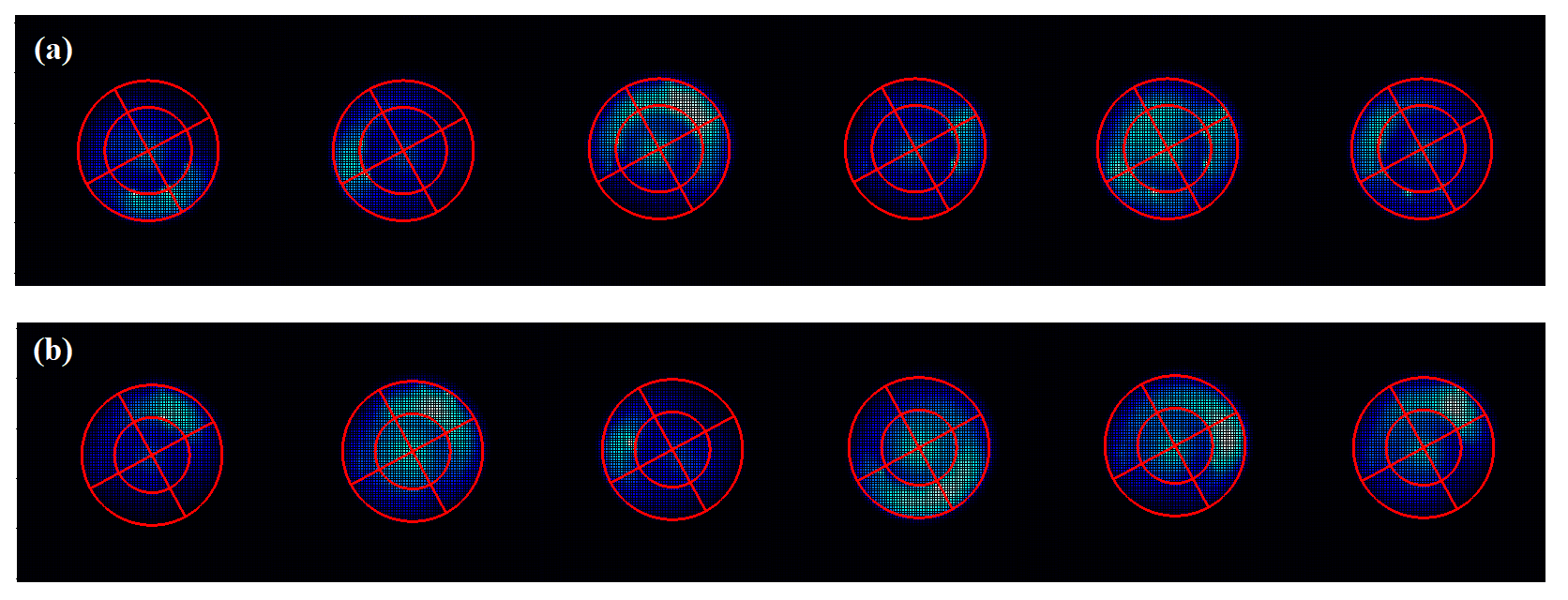}
	\caption{Fast camera images showing that azimuthal modes can be driven in either (a) $\mathbf{E}\times\mathbf{B}$ or (b) $-\mathbf{E}\times\mathbf{B}$ directions at \mbox{10 kHz}, which is different from the natural spoke frequency. From left to right, the consecutive camera images are taken at a time interval of $22\,\mu$s. The inner red circle marks the center pole of the cylindrical Hall thruster, whose channel wall is marked by the outer red circle. The four anode segments are located in the annulus regions between the red lines. The driven azimuthal modes are not a simple $m=1$ coherent mode.}
	\label{fig:image}
\end{figure*}

At conditions described above, an $m=1$ azimuthal mode naturally occurs in the CHT with frequency of about \mbox{$6$ kHz} in the passive operation mode. 
When active driving is turned on, the CHT plasma responds to the strong driving in the same direction at the same frequency. For example, when driven at \mbox{$10$ kHz} in the $\mathbf{E}\times\mathbf{B}$ direction, an $m=1$ azimuthal mode is induced in the same direction with the same frequency, as shown by the fast camera image in Fig.~\ref{fig:image}a. 
Similarly, when driven in the opposite direction (Fig.~\ref{fig:image}b), the induced azimuthal mode rotates in the $-\mathbf{E}\times\mathbf{B}$ direction at the driving frequency.
In Fig.~\ref{fig:image}, the false blue color corresponds to the intensity of the total light emission from the plasma. 
The inner red circle marks the center pole and the outer red circle marks the wall of the CHT channel. The anode segments are located in the annulus regions between the red lines.

\begin{figure}[b]
	\centering
	\includegraphics[width=0.5\textwidth]{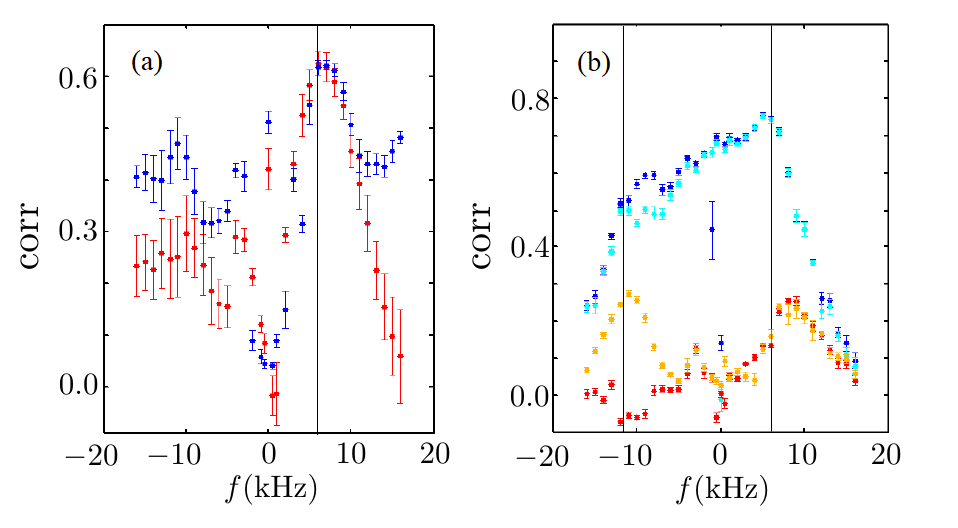}
	\caption{The coherence of the driven azimuthal modes can be measured using lag correlations of the image brightness (a) and the ion saturation currents (b) over the four anode segments. The axially integrated brightness is measured by a fast camera. When the maximum lag correlation (blue) is close to the lag-$\Delta T/4$ correlation (red), the $m=1$ mode dominates. This mode reaches maximum coherence when driven at the natural frequency \mbox{$f\approx6$ kHz}. 
	The ion saturation currents are measured by two sets of Langmuir probes. For the inner probes, the maximum lag correlation (dark blue) is always close to the lag-$\Delta T/4$ correlation (light blue), indicating that $m=1$ mode always dominates in the near-anode region. For outer probes, the maximum lag correlation (orange) deviates from the lag-$\Delta T/4$ correlation (red) for negative driving, which indicates that anode driving has less control over regions near the thruster exit.}
	\label{fig:lag}
\end{figure}

Although the plasma basically follows the anode forcing, the coherence of the azimuthal mode varies, depending on the frequency and direction of the driving. 
The coherence of the azimuthal mode can be measured using lag correlations between individual anode segments, where the lag correlation between two time series $A$ and $B$ is defined as $\textrm{corr}[A(t),B(t+\Delta t)]$. 
Using the fast camera images, the averaged brightness over each anode segment is computed. At each driving frequency, using the four brightness time series, six pairs of lag correlations are then computed. The average of the six lag correlations is represented by a colored dot and the spread is represented by the error bar in Fig.~\ref{fig:lag}a.  
For fixed $\Delta t=T/4$ between adjacent segments, the lag correlation (red) measures the coherence of the $m=1$ azimuthal mode. By allowing $\Delta t$ to vary, the lag correlation can be maximized (blue), which measures the coherence of the dominant azimuthal mode. 
At $f=0$, the mode is natural, and the coherence of the natural spoke $\sim 0.5$. When driven at low frequency \mbox{$f\sim \pm1$ kHz}, the azimuthal mode is suppressed. As the driving frequency approaches the natural frequency \mbox{$f=6$ kHz}, the azimuthal mode is enhanced. Driving at even higher frequencies excite other azimuthal modes, while the $m=1$ mode becomes suppressed. The brightness coherence depends less dramatically on the frequency when the mode is driven in the negative direction.

To see how the coherence of the azimuthal mode depends on the axial location along the thruster channel, data from the two sets of Langmuir probes, which measure ion saturation currents either in the ionization or the acceleration zone of the thruster, are also used to compute the lag correlations (Fig.~\ref{fig:lag}b).
For measurements taken in the ionization zone, the maximum lag correlation (dark blue) is close to the $\Delta t=T/4$ lag correlation (light blue), for both positive and negative driving frequencies. This indicates that the plasma response near the anode is always dominated by the driven $m=1$ azimuthal mode.
When $f>0$, the coherence of the mode slowly increases, until a maximum is reached at the natural spoke frequency \mbox{$f\approx6$ kHz}, beyond which the coherence drops rapidly. 
When $f<0$, the coherence slowly decreases until \mbox{$f\approx-12$ kHz}, below which the coherence drops sharply.
In comparison, for measurements taken in the acceleration zone, the coherence is significantly lower. This indicates that anode driving becomes less effective in the downstream of the CHT plasma.
This is expected not only because the downstream plasma is further away from the anode, but also because the downstream plasma naturally oscillates at a different azimuthal wave number \cite{Parker2010transition}.
In our driving experiments, the maximum lag correlation (orange) is close to the $\Delta t=T/4$ lag correlation (red) only for positive driving frequencies, indicating that the $m=1$ mode dominates only for driving in the $\mathbf{E}\times\mathbf{B}$ direction.
When $f>0$, the coherence peaks at a frequency slightly higher than the natural spoke frequency \mbox{$f\approx6$ kHz}. 
On the other hand, when $f<0$, the coherence reaches maximum near \mbox{$f=-12$ kHz}. The maximum lag correlation is attained for $\Delta t\approx -T/12$, indicating that near the thruster exist, the azimuthal mode is no longer the simple $m=1$ mode driven at the anode.

\begin{figure}[t]
	\centering
	\includegraphics[width=0.40\textwidth]{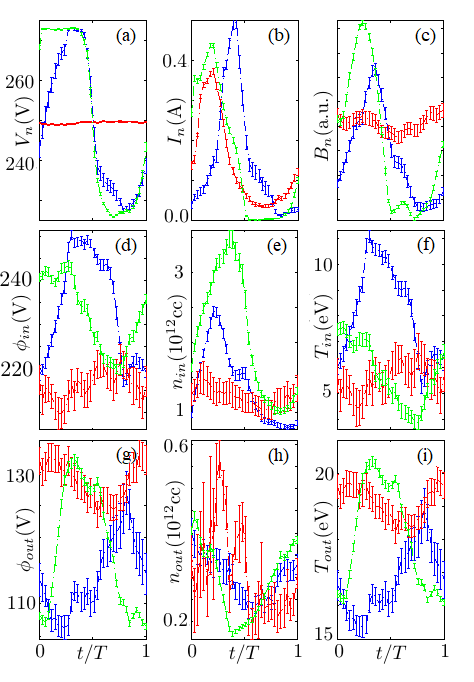}
	\caption{Plasma parameters for natural spoke (red), \mbox{$f=6$ kHz} driven spoke (green), and \mbox{$f=-6$ kHz} driven spoke (blue) are folded and averaged within one period $T$. (a) The actual voltage on an anode segment $V_n$ serves as a reference signal. (b) The discharge currents $I_n$ through the segment peak higher in the driven cases. (c) The brightness over the segment $B_n$ is roughly in phase with $I_n$. The plasma potential $\phi_{in}$ (d), density $n_{in}$ (e), and temperature $T_{in}$ (f) near the anode are measured by one set of Langmuir probes and emissive probes. Using another set of probes, the plasma potential $\phi_{out}$ (g), density $n_{out}$ (h), and temperature $T_{out}$ (i) are also measured near the thruster exist. The different phase relations unveil that azimuthal modes are different in the three cases, even though they have the same frequency.}
	\label{fig:para}
\end{figure}

The varying degree of coherence is correlated with different response of the plasma. For example, in Fig.~\ref{fig:para}, we compare plasma parameters in three cases: the natural mode without driving (red), the driven mode with \mbox{$f=6$ kHz} (green), and the mode driven in the opposite direction with \mbox{$f=-6$ kHz} (blue).
Time series of the measurements, recorded by a digital oscilloscope, are folded into a single period during post processing. The averages within the period are represented by the dots, and the standard deviations are represented by the error bars. 
As a reference signal, the actual voltage $V_n$ on an anode segment (Fig.~\ref{fig:para}a) is measured. 
The discharge current $I_n$ through the anode segment (Fig.~\ref{fig:para}b) strongly oscillates due to the azimuthal mode, while the total discharged current, summed over all four segments, varies only within $\sim10\%$.
The axially integrated brightness $B_n$ on the anode segment (Fig.~\ref{fig:para}c), as seen by the fast camera, is roughly in phase with $I_n$, so is the plasma density both in the ionization region ($n_{in}$, Fig.~\ref{fig:para}e) and in the acceleration region ($n_{out}$, Fig.~\ref{fig:para}h).
Roughly speaking, the plasma potential in the ionization region ($\phi_{in}$, Fig.~\ref{fig:para}d) and the electron temperature in the same region ($T_{in}$, Fig.~\ref{fig:para}f) are in phase, so are the plasma potential ($\phi_{out}$, Fig.~\ref{fig:para}g) and the electron temperature ($T_{out}$, Fig.~\ref{fig:para}i) in phase in the acceleration region. 
For the natural spoke (red), $\phi_{in}$ and $T_{in}$ are roughly out of phase with $I_n$, while $\phi_{out}$ and $T_{out}$ are roughly in phase with $I_n$.
In comparison, for \mbox{$f=6$ kHz} driving (green), $\phi_{in}$ and $T_{in}$ are roughly in phase with $I_n$, while $\phi_{out}$ and $T_{out}$ lag behind $I_n$ by $\sim T/4$.
Finally, for \mbox{$f=-6$ kHz} driving (blue), $\phi_{in}$ and $T_{in}$ are roughly in phase with $I_n$, while $\phi_{out}$ and $T_{out}$ lead ahead of $I_n$ by $\sim T/4$.
The different phase relations unveil that the physical processes beneath the azimuthal modes are very different in these three cases, even though the frequency of the modes are the same.

Although the plasma responds microscopically to the anode forcing, the averaged current utilization, propellant utilization, and plume angle of the CHT show little dependence on the driving frequency between \mbox{$-16$ kHz $<f<$ $16$ kHz}.
Using measurements from the plume probes, the current utilization, which is measured as the ratio of the ion current to the total discharge current, is found to be $\approx45\%$ and varies by only $2\%$ in response to driving at different frequencies.
Similarly, the propellant utilization, which is measured as the ratio of the ion mass flow rate to the anode mass flow rate 
is $\approx83\%$ and varies by only $3\%$.
Finally, the half-plume angle, which is defined as the angle containing $90\%$ of the total ion current, is $\approx67^\circ$ and varies by only $1\%$ in response to the driving. 
This suggests that the effect of anode driving on the overall performance is insignificant, when the CHT is operated in the aforementioned regime where the spoke already occurs naturally in the absence of the driving.

\section{Discussion and conclusion \label{sec:conclusion}}
In contrast to previous work where the azimuthal modes are induced or suppressed by changing discharge conditions, here we keep the discharge conditions fixed and control the azimuthal activities by only changing the anode boundary conditions. 
It is important to note that although we apply a rotating voltage on the segmented anode, the spatially averaged voltage on the entire anode is time independent and remains the same for all driving frequencies. This experimental setup enables us to distinguish correlations versus causal relations between the occurrence of the spoke and the performance change. 
While previous studies can only demonstrate that the performance of Hall plasma devices changes when the spoke appears, they cannot tell whether the performance change is due to the occurrence of the spoke, or both the performance change and the occurrence of the spoke are consequences of other changes in the operation conditions.
Results from our experiments now suggest that the later may be what happens in a CHT, with the caveat of the facility effects at SHTF. 
Beyond our experiments, which are conducted at moderate background pressure (50 microtorr), it is important to also investigate what happens at lower background pressure, at which the natural spoke is known to behave differently \cite{Raitses2010background}.
At the operation conditions investigated here, driving affects the spoke but does not change the overall performance, suggesting that the spoke is probably not causally related to performance changes.
Further measurements of voltage utilization and ion velocity distributions, for example, using time-resolved laser induced fluorescence \cite{Diallo2015time}, may confirm or refute the above conjecture.
Future studies analyzing effects of the background pressure, the driving voltage, and the operation conditions\footnote{Under conditions where the breathing mode naturally occurs, driving the entire anode noticeably changes the CHT performance \cite{Romadanov18}.} are also necessary for providing a more complete physical picture.


In summary, we demonstrate that azimuthal oscillations in a CHT plasma may be controlled by changing the boundary conditions on the anode. 
In particular, we show that by applying a rotating potential on a segmented anode, the coherence of the naturally occurring $m=1$ mode can either be enhanced or suppressed, depending on the direction and the frequency of driving.
The anode forcing induces local changes of plasma parameters, and the driven mode is physically distinct from the natural spoke.
The effects of driving are strongest in the near-anode region and decay in the downstream of the plasma. 
Although azimuthal diving has negligible effect on the averaged discharge characteristics of the CHT at operation conditions investigated here, local manipulations of plasma parameters may be useful for other Hall plasma devices.

\ack
This work was partially supported by AFOSR. We thank Ahmed Diallo and Stephane Mazouffre for fruitful discussion. We thank Scott Keller and Leland Ellison for help with experiments and Alexandre Merzhevskiy for expertise on electronics.

\bibliographystyle{iopart-num}
\providecommand{\newblock}{}

\end{document}